\documentclass[10pt]{amsart}
\usepackage{amsmath,amstext,amsbsy,amsopn,amsthm,%
amscd,amsfonts,amssymb}
\usepackage{bm}
\usepackage{epsfig}
\usepackage{graphpap}
\usepackage{mathrsfs}

\theoremstyle{definition}
\newtheorem{defn}{Definition}[section]
\theoremstyle{remark}
\newtheorem{remk}[defn]{Remark}
\newtheorem*{ack}{Acknowledgements}

\newcommand{\REMOVE}[1]{}

\newcommand{\eps}{\varepsilon}

\newcommand{\gO}{\varOmega}
\newcommand{\ofO}{{(\gO)}}
\newcommand{\ba}{\begin{align*}}
\newcommand{\ea}{{\end{align*}}}
\newcommand{\les}{\leqslant}

\newcommand{\dee}{\mathrm{d}}
\newcommand{\tto}{\longrightarrow}
\newcommand{\abs}[1]{\left|#1\right|}
\newcommand{\comp}{\raisebox{0.75pt}{\mbox{$\:{\scriptstyle \circ}\:$}}}
\newcommand{\cl}[1]{{\overline{#1}}}

\newcommand{\cee}[1]{C^{#1}\!}
\newcommand{\lone}{{L^1}}
\newcommand{\linf}{{L^\infty}}
\newcommand{\lomu}{{L^{1,\mu}}}
\newcommand{\limu}{{L^{\infty,\mu}}}
\newcommand{\sob}{{W^1_1}}
\newcommand{\sobb}{\sob(\gO,\rthree)}
\newcommand{\lomub}{\lomu(\cl\gO,\rthree)}
\newcommand{\rigs}{\mathscr{R}}
\newcommand{\qrigs}{\!/\!\rigs}
\newcommand{\ld}{{LD}}
\newcommand{\ldo}{\ld\ofO}
\newcommand{\trace}{\gamma}
\newcommand{\rig}{r}
\newcommand{\anv}{\omega}
\newcommand{\I}{I}
\newcommand{\pr}{{\proj_{\rigs}}}
\newcommand{\po}{{\proj_{0}}}
\newcommand{\normp}[1]{{\norm{#1}'_\ld}}

\newcommand{\rsix}{{\reals^{6}}}

\newcommand{\bndb}{{\partial\body}}
\newcommand{\bndo}{{\partial\gO}}
\newcommand{\cloo}{{\cl\gO}}

\newcommand{\reals}{\ensuremath{\mathbb{R}}}
\newcommand{\rthree}{\ensuremath{\reals^{\scriptscriptstyle 3}}}

\DeclareMathOperator{\ess}{ess}
\DeclareMathOperator*{\esssup}{ess\;sup}
\DeclareMathOperator{\kernel}{Kernel}
\newcommand{\from}{\colon}
\DeclareMathOperator{\image}{Image}
\newcommand{\incl}{\iota}
\newcommand{\proj}{\pi}
\newcommand{\by}{\!/\!}
\newcommand{\vs}{\ensuremath{\mathbf{W}}}

\newcommand{\ino}[1]{\int\limits_{#1}}
\newcommand{\dby}[2]{\frac{\partial #1}{\partial #2}}
\newcommand{\half}{\ensuremath{{\frac{1}{2}}}}

\newcommand{\extnd}{\ensuremath{\delta}}
\newcommand{\paren}[1]{\left(#1\right)}

\newcommand{\braces}[1]{\left\{#1\right\}}
\newcommand{\norm}[1]{\left\|#1\right\|}
\newcommand{\resto}[1]{\raisebox{-1.2pt}{$\big\vert$}_{#1}}
\newcommand{\pis}{\ensuremath{x}}

\newcommand{\body}{\ensuremath{B}}%

\newcommand{\dV}{\dee{V}}
\newcommand{\dA}{\dee{A}}

\newcommand{\vf}{w}

\newcommand{\fc}{F}
\newcommand{\bfc}{{b}}
\newcommand{\sfc}{{t}}
\newcommand{\st}{\sigma}

\newcommand{\stm}{S}
\begin{document}
\title[Stress Concentrations and Equilibrium -- \today]{Generalized Stress Concentration Factors
 for Equilibrated Forces and Stresses}

\author{Reuven Segev}%

\address{Department of Mechanical Engineering,
Ben-Gurion University\\ P.O.Box~653, Beer-Sheva 84105
Israel}

\email{rsegev@bgu.ac.il}
\date{\today\\
Department of Mechanical Engineering,
Ben-Gurion University, Beer Sheva 84105, Israel,\\
E-mail: rsegev@bgu.ac.il, Fax:
972-8-6472814, Phone: 972-8-6477043}

\begin{abstract}
As a sequel to a recent work we consider the
generalized stress concentration factor, a
purely geometric property of a body that for
the various loadings, indicates the ratio
between the maximum of the optimal stress
and maximum of the loading fields. The
optimal stress concentration factor pertains
to a stress field that satisfies the
principle of virtual work and for which the
stress concentration factor is minimal.
Unlike the previous work, we require that
the external loading be equilibrated and
that the stress field be a symmetric tensor
field.
%
  \\[2pt]
\end{abstract}

\maketitle
\REMOVE{
\begin{flushright}
\REMOVE{
\textit{%
In memory of my friends
}\hspace*{3cm}\\{\textsc{
{Isaac Feldman, 1954--1973,\\
Amir Moses, 1954--1996,\\
Ilan Ramon (Wolferman), 1954--2003.}}}\\
\textit{%
} }
{
\aalcuin
In memory of my friends\hspace*{3.5cm}} \\
{\alcuin ISAAC FELDMAN, 1954--1973,\\
AMIR MOSES, 1954--1996,\\
ILAN RAMON (WOLFERMAN), 1954--2003. }
\end{flushright}
\vspace*{0.3cm}
}
\thispagestyle {empty}
\section{{Introduction}}
In a recent article \cite{SCF} we introduced
the notion of a generalized stress
concentration factor as a quantitative
measure of how bad is the geometry of a body
in terms of the ratio between the maximal
stresses and the maximum of the applied
loads. Specifically, generalized stress
concentration factors may be described as
follows. Let $\fc$ be a force on a body
$\gO$ that is given in terms of a body force
field $\bfc$ and a surface force field
$\sfc$ and let $\st$ be any stress field
that is in equilibrium with $\fc$. Then, the
stress concentration factor for the pair
$\fc$, $\st$ is given by
\[
K_{\fc,\st}=
 \frac{\sup_{x}\braces{\abs{\st(x)}}}
      {\sup_{x,y}\braces{\abs{\bfc(x)},\abs{\sfc(y)}}}\,,
      \quad
      x\in\gO,\; y\in\bndo.
\]
Here, for $\abs{\st(\pis)}$ we use some norm
$\abs\cdot$ on the space of stresses at a
point---a finite dimensional space.  We note
that since we do not specify a constitutive
relation, for each force $\fc$ there is a
class $\Sigma_\fc$ of stress fields $\st$
that are in equilibrium with $\fc$.

Next, the optimal stress concentration
factor for the force $\fc$ is defined by
\[
K_\fc=\inf_{\st\in\Sigma_\fc}\braces{K_{\fc,\st}},
\]
i.e., it is the least stress concentration
factor when we allow the stress field to
vary over all fields that are in equilibrium
with $\fc$. Finally, the generalized stress
concentration factor $K$---a purely
geometric property of $\gO$---is defined by
\[
K=\sup_\fc\braces{K_\fc}
    =\sup_\fc\frac{\inf_{\st\in\Sigma_\fc}
          \braces{\sup_{x}\braces{\abs{\st(x)}}}}
              {\sup_{x,y}\braces{\abs{\bfc(x)},\abs{\sfc(y)}}},
\]
where $\fc$ varies over all forces that may
be applied to the body.  Thus, the
generalized stress concentration factor
reflects the worst case of loading of the
body.

It was shown in \cite{SCF} that the
generalized stress concentration factor is
equal to the norm of a mapping associated
with the trace operator of Sobolev mappings.
Specifically, it was shown that when suprema
in the expressions above are replaced by
essential suprema, then,
\[
K= \sup_{\phi\in\sobb}
   \frac{\ino\gO\abs{\phi}\,\dV
             +\ino\bndo\abs{\hat\phi}\,\dA}
        {\ino\gO\abs{\phi}\,\dV+
           \ino\gO\abs{\nabla\phi}\,\dV}\,,
\]
where $\sobb$ is the Sobolev space of
integrable vector fields $\phi$ on $\gO$
whose gradients $\nabla\phi$ are also
integrable, and $\hat\phi$ is the trace of
$\phi\in\sobb$ on $\bndo$ (whose existence
is a basic property of Sobolev spaces).

Consider the Radon measure $\mu$ on $\cl\gO$
defined by
\[
\mu(D)=V(D\cap\gO)+A(D\cap\bndo)
\]
($V$ and $A$ are the volume and area
measures, respectively), and let $\lomub$ be
the space of fields on $\cloo$ that are
integrable relative to $\mu$ equipped with
the $\lomu$-norm so
\[
\norm\vf_\lomu=\ino\gO\abs\vf\,\dV
     +\ino\bndo\abs\vf\,\dA.
\]
Then, the trace operator induces an
extension mapping
$\delta\from\sobb\to\lomub$ and the
expression for the generalized stress
concentration factor above may be written in
the form
\[
K=\norm\extnd
\]
---the basic result of \cite{SCF}.

The treatment in \cite{SCF} allows stresses
and forces that are more general than those
treated usually in continuum mechanics. In
addition to the usual stress tensor
$\st_{im}$ the stress object contains a self
force field $\st_i$.  Furthermore, the
stress field need not be symmetric and the
resultants and total torques due to the
forces $\fc$ need not vanish.  The
generalized form of the equilibrium
equations between the forces and stresses
was taken in the form
\[
\ino\gO\bfc_i\vf_i\,\dV
            +\ino\bndo\sfc_i\vf_i\,\dA
=\ino\gO\st_{i}\vf_i\,\dV
 + \ino\gO\st_{ik}\vf_{i,k}\,\dV.
\]
Thus, the infimum in the definition of the
optimal stress concentration factor may be
attained for a stress field that is not
admissible physically.

In the present work we restrict the
admissible stress fields to symmetric tensor
fields and the forces are required to have
zero resultants and total torques.  These
requirements are well known to be equivalent
to the requirements that the power produced
by the forces and stresses on rigid velocity
fields vanishes.

The expression for the generalized stress
concentration factor we obtain here for the
rigid velocity invariant forces and stresses
may be written as
\[
K=\norm{\extnd\qrigs},
\]
where $\rigs$ denotes the collection of
rigid velocity fields, a subspace of the
function-spaces we are considering. The
extension mapping
\[
\extnd\qrigs\from\ldo\qrigs
        \tto\lomu(\cl\gO,\rthree)\qrigs
\]
between the corresponding quotient spaces is
given by
$\extnd\qrigs([\vf])=[\extnd(\vf)]$.  It is
well defined for elements of the space
$\ldo$ containing the vector fields $\vf$ of
integrable stretchings
$$
\eps(\vf)=\half(\nabla\vf+(\nabla\vf)^T).
$$
The space $\ldo$ and its properties (see
\cite{Temam85,TemamStrang,Strauss,Temam81,Ambrosio,Ebobisse},
and \cite{Kohn82} for nonlinear strains) are
the main technical tools we use in this
work.


For a projection mapping that gives an
approximating rigid velocity field to any
vector field $\vf$ and a corresponding
$\vf_0$ that has zero rigid component, this
result may be written more specifically as
\[
K=\norm{\delta_0}
  =\sup_{\vf_0\in\ldo_0}\frac
  {\inf_{\rig\in\rigs}\braces{
     \displaystyle\int_\gO\sum_i\abs{\vf_{0i}-\rig_i}\,\dV
             +\displaystyle{\int}_{\!\!\!\partial\gO}\sum_i\abs{\vf_{0i}-\rig_i}\,\dA}}
     {\half\displaystyle\int_\gO\sum_{i,m}\abs{\vf_{0i,m}+\vf_{0m,i}}\,\dV}\,.
\]
Here $\extnd_0$ is the extension mapping for
vector fields having zero rigid components
and $\ldo_0$ is the space of vector fields
in $\ldo$ having zero rigid components.

Section 2 presents some properties of rigid
velocity fields, stretchings and the
approximations of velocity fields by rigid
ones.  Section 3 outlines the definitions
and results pertaining to the space $\ldo$
and is based on \cite{Temam85}.  Finally,
Section 4 applies the properties of
$\ld$-fields to the problem under
consideration.  Some details regarding the
notation we use and results on normed spaces
and their normed dual spaces are available
in \cite{SCF}.

\vspace*{2mm}
I wish to thank R.~Kohn for pointing the
$BD$-literature to me and F.~Ebobisse for
his Ph.D.~thesis and comments on it.

\section{Preliminaries on Stretchings and Rigid Velocities}

\subsection{Basic definitions}
Let $\gO$ be an open and bounded
3-dimensional submanifold of $\rthree$ with
volume $\abs\gO$ having a differentiable
boundary and $\vf$ a vector field over
$\gO$.  We set $\eps(\vf)$ to be the tensor
field
\[
\eps(\vf)_{im}=\half(\vf_{i,m}+\vf_{m,i}),
\]
i.e., the symmetric part of the gradient. As
$\vf$ is interpreted physically as a
velocity field over the body, $\eps(\vf)$ is
interpreted as the stretching.
Alternatively, if $\vf$ is interpreted as an
infinitesimal displacement field,
$\eps(\vf)$ is the corresponding linear
strain. In the sequel we will refer to
$\eps(\vf)$ as the \emph{stretching}
associated with $\vf$. Here, the partial
derivatives are interpreted as the
distributional derivatives so one need not
care about the regularity of $\vf$.

We identify the space of symmetric
$3\times3$ matrices with $\rsix$.  For a
symmetric tensor field $\eps$ whose
components are integrable functions we use
the $\lone$-norm
\[
\norm\eps=\sum_{i,m}\norm{\eps_{im}}_\lone.
\]
This norm maybe replaced by other equivalent
norms (possibly norms invariant under
coordinate transformations).  Thus, the
space of $\lone$-stretching fields is
represented by $\lone(\gO,\rsix)$ with the
$\lone$-norm as above.

A vector field $\vf$ on $\gO$ is \emph{of
integrable stretching} if its components are
integrable and if each component
$\eps(\vf)_{im}\in\lone\ofO$. It can be
shown that this definition is coordinate
independent.  The vector space of velocity
fields having integrable stretchings will be
denoted by $\ldo$. This space is normed by
\[
\norm\vf_\ld=\sum_i\norm{\vf_i}_\lone+
            \sum_{i,m}\norm{\eps(\vf)_{im}}_\lone.
\]
Clearly, we have a continuous linear
inclusion
$\ld(\gO,\rthree)\tto\lone(\gO,\rthree)$. In
addition, $\vf\mapsto\eps(\vf)$ is given by
a continuous linear mapping
\[
\eps\from\ldo\tto\lone(\gO,\rsix).
\]

\subsection{The subspace of rigid velocities}
 \label{RigidVel}
A \emph{rigid} velocity (or displacement)
field is of the form
\[
\vf(x)=a+{\anv}\times{\pis},\quad \pis\in\gO
\]
where $a$ and ${\anv}$ are fixed in
$\rthree$ and ${\anv}\times{\pis}$ is the
vector product. We can replace
${\anv}\times{\pis}$ with $\tilde\anv(\pis)$
where $\tilde\anv$ is the associated skew
symmetric matrix so
$\vf(\pis)=a+\tilde\anv(\pis)$.  We will
denote the 6-dimensional space of rigid body
velocities by $\rigs$.  For a rigid motion
\[
\tilde\anv_{im}=\half(\vf_{i,m}-\vf_{m,i}),
\]
an expression that is extended to the
non-rigid situation and defines the
vorticity vector field so
$\vf_{i,m}=\eps(\vf)_{im}+\tilde\anv_{im}$.

Considering the kernel of the stretching
mapping
\( \eps\from\ldo\tto\lone(\gO,\rsix), \) a
theorem whose classical version is due to
Liouville states (see
\cite[pp.~18--19]{Temam85}) that
$\kernel\eps=\rigs$.

\subsection{Approximation by rigid velocities}
 \label{Approx}

We now wish to consider the approximation of
a velocity field by a rigid velocity field.
Let $\rho$ be a Radon measure on $\cl\gO$
and $1\les{p}\les\infty$. For a given
$\vf\in{L}^{p,\rho}(\cl\gO,\rthree)$, we
wish to find the rigid velocity $\rig$ for
\[
\inf_{\rig'\in\rigs}\paren{\norm{\vf-\rig'}_{L^{p,\rho}}}^p
=\inf_{\rig'\in\rigs}
  \ino{\cl\gO}\sum_i{\abs{\vf_i-\rig'_i}^p}\,\dee\rho
\]
is attained. Thus we are looking for vectors
$a$ and $b$ that minimize
\[
e=\ino{\cl\gO}\sum_i{\abs{\vf_i-a_i-\eps_{ijk}{b_jx_k}}^p}\dee\rho.
\]
We have
\begin{align*}
\dby{e}{a_l}&=
 \ino{\cl\gO}p\sum_i\abs{\vf_i-a_i-\eps_{ijk}b_jx_k}^{p-1}
        \frac{\paren{\vf_i-a_i-\eps_{ijk}b_jx_k}}{{\abs{\vf_i-a_i-\eps_{ijk}b_jx_k}}}
               (-\delta_{il})\,\dee\rho,
           \\
           \dby{e}{b_l}&=
 \ino{\cl\gO}p\sum_i\abs{\vf_i-a_i-\eps_{ijk}b_jx_k}^{p-1}
        \frac{\paren{\vf_i-a_i-\eps_{ijk}b_jx_k}}{{\abs{\vf_i-a_i-\eps_{ijk}b_jx_k}}}
               (-\eps_{ijk}\delta_{jl}x_k)\,\dee\rho,
           \end{align*}
and we obtain the 6 equations for the
minimum with the 6 unknowns $a_l$, $b_m$
\begin{align*}
0&=
  \ino{\cl\gO}\abs{\vf_l-a_l-\eps_{ljk}b_jx_k}^{p-2}
        {\paren{\vf_l-a_l-\eps_{ljk}b_jx_k}}\,\dee\rho,
           \\0&=
  \ino{\cl\gO}\sum_i\abs{\vf_i-a_i-\eps_{ijk}b_jx_k}^{p-2}
        {\paren{\vf_i-a_i-\eps_{ijk}b_jx_k}}\eps_{ilk}x_k\,\dee\rho.
\end{align*}

Particularly simple are the equations for
$p=2$.  In this case we obtain
\begin{align*}
\ino{\cl\gO}\vf\,\dee\rho=\ino{\cl\gO}\rig\,\dee\rho,
\quad\text{and}\quad
 \ino{\cl\gO}x\times\vf\,\dee\rho=
         \ino{\cl\gO}x\times\rig\,\dee\rho.
\end{align*}
If we interpret $\rho$ as a mass
distribution on $\cl\gO$, these two
conditions simply state that the best rigid
velocity approximations should give the same
momentum and angular momentum as the
original field.

Of particular interest (see
\cite[p.~120]{Temam85}) is the case where
$\rho$ is the volume measure on $\gO$. Set
$\bar\pis$ to be the center of volume of
$\gO$, i.e.,
\[
\bar\pis=\frac{1}{\abs\gO}\ino\gO\pis\,\dV.
\]
Without loss of generality we will assume
that $\bar\pis=0$ (for else we may replace
$\pis$ by $\pis-\bar\pis$ in the sequel).

 Let $\bar\vf$ be
the mean of the field $\vf$ and $\I$ the
inertia matrix relative to the center of
volume, so
\[
 \bar\vf=\frac{1}{\abs\gO}\ino\gO\vf\,\dV,
 \quad
 \I_{im}=\ino\gO(\pis_k\pis_k\delta_{im}-\pis_i\pis_m)\,\dV
\]
and
\[
\I(\anv)=\ino\gO\pis\times(\anv\times\pis)\,\dV.
\]
The inertia matrix is symmetric and positive
definite and so the solution for $\rig$
gives
\[
\rig=\bar\vf+\anv\times\pis
\]
with $\bar\vf$ as above and
\[
\anv=\I^{-1}\left(\ino\gO\pis\times\vf\,\dV\right).
\]
Thus, $\vf\mapsto(\bar\vf+\anv\times\pis)$,
with $\bar\vf$ and $\anv$ as above, is well
defined for integrable velocity fields and
we obtain a mapping
\[
\proj_\rigs\from\lone(\gO,\rthree)\tto\rigs.
\]
It is straightforward to show that $\pr$ is
indeed a linear projection onto $\rigs$.

Also of interest below will be the case
where $p=1$ and and the measure $\rho$ is
given by
\[
\rho(D)=\mu(D)=V(D\cap\gO)+A(D\cap\partial\gO),
\]
as in Section~1.  The conditions for best
approximations $\rig=a+b\times\pis$ assume
the form
\[
  \ino{\gO}\frac{\paren{\vf_l-a_l-\eps_{ljk}b_jx_k}}
        {\abs{\vf_l-a_l-\eps_{ljk}b_jx_k}}\,\dV
  +\ino{\partial\gO}\frac{\paren{\vf_l-a_l-\eps_{ljk}b_jx_k}}
        {\abs{\vf_l-a_l-\eps_{ljk}b_jx_k}}\,\dA=0 ,
\]
\[
  \ino{\gO}\sum_i\frac{\paren{\vf_i-a_i-\eps_{ijk}b_jx_k}}
        {\abs{\vf_i-a_i-\eps_{ijk}b_jx_k}}\,\eps_{ilk}x_k\,\dV
  + \ino{\partial\gO}\sum_i\frac{\paren{\vf_i-a_i-\eps_{ijk}b_jx_k}}
        {\abs{\vf_i-a_i-\eps_{ijk}b_jx_k}}\,\eps_{ilk}x_k\,\dA=0,
\]
where $z/\abs{z}$ is taken as 0 for $z=0$.
(For an analysis of $L^1$-approximations see
\cite{Pinkus} and reference cited therein.)

\subsection{Distortions}
 \label{Distortions}
Let $\vs$ be a vector space of velocities on
$\gO$ containing the rigid velocities
$\rigs$ and let $\vf_1$ and $\vf_2$ be two
velocity fields in $\vs$. We will say that
the two have the same \emph{distortion} if
$\vf_2=\vf_1+\rig$ for some rigid motion
$\rig\in\rigs$.  This clearly generates an
equivalence relation on $\vs$ and the
corresponding quotient space $\vs\qrigs$
will be referred to as the space of
distortions.  If $\chi$ is an element of
$\vs\qrigs$ then $\eps(\vf)$ is the same for
all members of $\vf\in\chi$.  The natural
projection
\[
\proj\from\vs\tto\vs\qrigs
\]
associates with each element $\vf\in\vs$ its
equivalence class
$[\vf]=\braces{\vf+\rig|\rig\in\rigs}$.

If $\vf$ is a normed space, then, the
induced norm on $\vf\qrigs$ is given by (see
Appendix \ref{AppQuot})
\[
\norm{[\vf]}=\inf_{\vf'\in[\vf]}\norm{\vf'}
             =\inf_{\rig\in\rigs}\norm{\vf-\rig}.
\]
Thus, the evaluation of the norm of a
distortion, is given by the best
approximation by a rigid velocity as
described above.

Let $\vs$ be a vector space of velocities
contained in $\lone(\gO,\rthree)$, then,
$\pr$ defined above induces an additional
projection
\[
\po(\vf)=\vf-\pr(\vf).
\]
The image of $\po$ is the kernel $\vs_0$ of
$\pr$ and it is the subspace of $\vs$
containing velocity fields having zero
approximating rigid velocities.  Clearly, we
have a bijection
$\beta\from\vs\qrigs\to\vs_0$. On $\vs_0$ we
have two equivalent norms: the norm it has
as a subspace of $\vs$ and the norm that
makes the bijection
$\beta\from\vs\qrigs\to\vs_0$ an isometry.

With the projections $\po$ and $\pr$, $\vs$
has a Whitney sum structure
$\vs=\vs_0\oplus\rigs$.

\subsection{Equilibrated forces}
Let $\vs$ be a vector space of velocities
(we assume that it contains the rigid
velocities). A force $\fc\in\vs^*$ is
equilibrated if $\fc(\rig)=0$ for all
$\rig\in\rigs$.  This is of course
equivalent to $\fc(\vf)=\fc(\vf+\rig)$ for
all $\rig\in\rigs$ so $\fc$ induces a unique
element of $\paren{\vs\qrigs}^*$.
Conversely, any element of
$G\in\paren{\vs\qrigs}^*$ induces an
equilibrated force $\fc$ by
$\fc(\vf)=G([\vf])$, where $[\vf]$ is the
equivalence class of $\vf$.  In other words,
as the quotient projection is surjective,
the dual mapping
$\proj^*\from(\vs\qrigs)^*\to\vs^*$ is
injective and its image---the collection of
equilibrated forces---is orthogonal to the
kernel of $\proj$.  Furthermore, as in
Appendix \ref{AppQuot}, $\proj^*$ is norm
preserving.
 Thus, we may
identify the collection of equilibrated
forces in $\vs^*$ with
$\paren{\vs\qrigs}^*$.

If $\incl_\rigs\from\rigs\to\vs$ is the
inclusion of the rigid velocities, then,
\[
\incl_\rigs^*\from\vs^*\tto\rigs^*,
\]
is a continuous and surjective mapping.  The
image $\incl_\rigs^*(\fc)$ will be referred
to as the total of the force. In particular,
its component dual to $\bar\vf$ will be
referred to as the force resultant and the
component dual to $\anv$ will be referred to
as the resultant torque. Thus, in
particular, the resultant force and torque
vanish for an equilibrated force. This
structure may be illustrated by the
sequences%
\[
\begin{CD}
0  @>>> \rigs  @>{\incl_\rigs}>> \vs
               @>\proj>>  \vs\qrigs @>>>
               0\phantom{.}
               \\
0  @<<< \rigs^*  @<{\incl^*_\rigs}<< \vs^*
              @<\proj^*<< (\vs\qrigs)^* @<<< 0.
\end{CD}
\]

Using the projection $\pr$ and the Whitney
sum structure it induces we have a Whitney
sum structure $\vs^*={\vs_0^*}\oplus\rigs^*$
and it is noted that the norm on $\vs_0^*$
is implied by the choice of norm on $\vs_0$.

\section{Fields of Integrable Stretchings}
In this section we list the basic properties
of vector fields of integrable stretching
(or deformation) as in \cite{Temam85} (see
also
\cite{TemamStrang,Strauss,Temam81,Ambrosio,Ebobisse}
and \cite{Kohn82} for nonlinear strains).
The presentation below is adapted to the
application we consider and is not
necessarily the most general.

  If both
$\vf$ and $\eps(\vf)$ are in $L^p$ for
$1<{p}<\infty$, the Korn inequality (see
\cite{Friedrichs}) implies that
$\vf\in\sob(\gO)$.  This would imply in
particular that $\vf$ has a trace on the
boundary of $\gO$.  However, as shown by
Ornstein \cite{Ornstein}, $\vf$ need not
necessarily be in $\sobb$ for the critical
value $p=1$. Nevertheless, the theory of
integrable stretchings shows that the trace
is well defined even for $p=1$.

\subsection{Definition}\label{Definition}
We recall that $\ldo$ is the vector space of
fields with integrable stretchings.  With
the norm
\[
\norm\vf_\ld=\sum_i\norm{\vf_i}_\lone+
            \sum_{i,m}\norm{\eps(\vf)_{im}}_\lone
\]
$\ldo$ is a Banach space.

\subsection{Approximation}
$\cee\infty(\cl\gO,\rthree)$ is dense in
$\ldo$.

\subsection{Traces}
 \label{Trace}
The trace operator can be extended from
$\sob(\gO,\rthree)$ onto $\ld(\gO,\rthree)$.
Thus, there is a unique continuous linear
mapping
\[
\trace\from\ld(\gO)\tto\lone(\partial\gO,\rthree)
\]
such that
$\trace(\vf)=\vf\resto{\partial\gO}$, for
every field $\vf$ of bounded stretching that
is a restriction to $\gO$ of a continuous
field on the closure $\cl\gO$. Thus, the
norm of the trace mapping is given by
\[
\norm\trace=\sup_{\vf\in\ld(\gO)}
           \frac{\norm{\trace(\vf)}_\lone}
             {\norm\vf_{\ld}}.
\]
As a result of the approximation of fields
of bounded stretchings by smooth vector
fields on $\cl\gO$, $\norm\trace$ may be
evaluated using smooth vector fields in the
expression above, i.e.,
\[
\norm{\trace}=\sup_{\vf\in\cee\infty(\cl\gO,\rthree)}
           \frac{\norm{\vf\resto{\partial\gO}}_\lone}
             {\norm\vf_{\ld}}\;.
\]

\subsection{Extensions}\label{Exten}
There is a continuous linear extension
operator%
$$
E\from\ldo\tto\ld(\rthree)
$$
such that $E(\vf)(\pis)=\vf(\pis)$ for
almost all $\pis\in\gO$.
\subsection{Regularity}
If $\vf$ is any distribution on $\gO$ whose
corresponding stretching is $\lone$, then
$\vf\in\lone(\gO,\rthree)$.

\subsection{Distortions of integrable stretching}
 \label{EquivNormQ}
On the space of $\ld$-distortions,
$\ldo\qrigs$, we have a natural norm
\[
\norm\chi=\inf_{\vf\in\chi}\norm{\vf}_\ld.
\]
This norm is equivalent to
\[
\norm{\eps(\chi)}=\sum_{i,m}\norm{\eps(\vf)_{im}}_\lone
\]
where $\vf$ is any member of $\chi$.
Clearly, the value of this expression is the
same for all members $\vf\in\chi$ and we can
use any other equivalent norm
 on the space of symmetric tensor fields.

Using the projection $\pr$ as above we
denote by $\ldo_0$  the kernel of $\pr$ and
by $\po$ the projection onto $\ldo_0$ so
$$
 (\po,\pr)\from\ldo
  \tto\ldo_0\oplus\rigs.
$$
Then, there is a
constant $C$ depending only on $\gO$ such
that
\[
\norm{\po(\vf)}_\lone=
 \norm{\vf-\pr(\vf)}_\lone\les
  C\norm{\eps(\vf)}_\lone.
\]
\subsection{Equivalent norms}
 \label{EquivNormLD}
Let $p$ be a continuous seminorm on $\ldo$
which is a norm on $\rigs$.  Then,
\[
p(\vf)+\norm{\eps(\vf)}_\lone
\]
is a norm on $\ldo$ which is equivalent to
the original norm in \ref{Definition}.


\section{Application to Equilibrated Forces and Stresses}

\subsection{$\ld$-velocity fields and forces}
The central object we consider is $\ldo$
whose elements are referred to as
$\ld$-velocity fields.  Elements of the dual
space $\ldo^*$ will be referred to as
$\ld$-forces.  Our objective is to represent
$\ld$-forces by stresses and by pairs
containing body forces and surface forces.

Rather than the original norm of
\ref{Definition} it will be convenient to
use an equivalent norm as asserted by
\ref{EquivNormLD} as follows.  Let
\[
\pr\from\ldo\tto\rigs
\]
 be the continuous linear projection defined
 in Paragraph \ref{Approx} and let
 $q\from\rigs\to\reals$, be a norm on the
 finite dimensional $\rigs$.  Then,
 \[
p=q\comp\pr\from\ldo\tto\reals
 \]
  is a continuous seminorm that is a norm on
$\rigs\subset\ldo$.  It follows from
\ref{EquivNormLD} that
\[
\normp\vf=q\paren{\pr(\vf)}+\norm{\eps(\vf)}_\lone
\]
is a norm on $\ldo$ which is equivalent to
the original norm defined in
\ref{Definition}.
\subsection{$\ld$-distortions}
With the norm $\normp\cdot$, the induced
norm on $\ldo\qrigs$ is given by
\[
\normp{[\vf]}=\inf_{\rig\in\rigs}\normp{\vf+\rig},
\]
so using $\pr(r)=r$, $\eps(\rig)=0$ and
choosing $\rig=-\pr(\vf)$, we have
\begin{align*}
\normp{[\vf]}
       &=\inf_{\rig\in\rigs}\braces{
             q(\pr(\vf+\rig))+\norm{\eps(\vf+\rig)}_\lone}
       \\&=\inf_{\rig\in\rigs}\braces{
             q(\pr(\vf)+\rig)+\norm{\eps(\vf)}_\lone}
       \\&=\norm{\eps(\vf)}_\lone.
\end{align*}

Let $\po\from\ldo\to\ldo_0$ be the
projection onto $\ldo_0\subset\ldo$, the
kernel of $\pr$. Then,
\begin{align*}
\norm{\po(\vf)}'_\ld
 &=\norm{\vf-\pr(\vf)}'_\ld
  \\&=q(\pr(\vf-\pr(\vf)))+\norm{\eps(\vf-\pr(\vf))}_\lone
  \\&=\norm{\eps(\vf)}_\lone.
\end{align*}

 We conclude that with our choice of norm
$\normp\cdot$ on $\ldo$, the two norms in
\ref{EquivNormQ} are not only equivalent but
are actually equal.  Thus, this choice makes
$\ldo_0$ isometrically isomorphic to
$\ldo\qrigs$.

\subsection{Equilibrated $\ld$-forces and their
representations
 by stresses}
Summarizing the results of the previous
sections we can draw the commutative diagram
\[
\begin{CD}
                 \ld(\gO)
                   @>\eps>> \lone(\gO,\rsix)
                               \\
 {}         @VV{\proj}V       {}   @|\\
                 \ld(\gO)\qrigs
                   @>\eps\qrigs>>
                             \lone(\gO,\rsix).
\end{CD}
\]
Here, Liouville's rigidity theorem implies
that the kernels of $\eps$ and $\proj$ are
identical, the rigid velocity fields, and
$\eps\qrigs$ given by
$\eps\qrigs(\chi)=\eps(\vf)$, for some
$\vf\in\chi$, is an isometric injection.

This allows us to represent
$\ld$-forces---elements of $\ldo^*$---using
the dual diagram.
\[
\begin{CD}
                 \ld(\gO)^*
                   @<\eps^*<<
                   \linf(\gO,\rsix)
                               \\
 {}         @AA{\proj^*}A       {}   @|\\
                 \paren{\ld(\gO)\qrigs}^*
                   @<(\eps\qrigs)^*<<
                             \linf(\gO,\rsix).
\end{CD}
\]
Now, $(\eps\qrigs)^*$ is surjective and as
in \cite{SCF} the Hahn-Banach Theorem
implies that any $T\in\paren{\ldo\qrigs}^*$
may be represented in the form
\[
T=(\eps\qrigs)^*(\st)
\]
for some essentially bounded symmetric
stress tensor field
$\st\in\linf(\gO,\rsix)$.  Furthermore, the
dual norm of $T$ is given by
\[
\norm{T}=\inf_{T=(\eps\qrigs)^*(\st)}\norm\st_\linf
 =\inf_{T=(\eps\qrigs)^*(\st)}\braces{
    \ess\sup_{i,m,\pis\in\gO}\abs{\st_{im}(\pis)}
  }.
\]
As $\proj^*$ is norm preserving (see
Appendix \ref{AppQuot}), the same holds for
any equilibrated $\ld$-force.  That is,
using the same argument for $(\ldo\qrigs)^*$
and the fact that $\proj^*$ is a
norm-preserving injection, any equilibrated
$\ld$-force $\stm\in\ldo^*$ may be
represented in the form
\[
\stm=\eps^*(\st)
\]
for some stress field $\st$ and
\[
\norm{\stm}=\inf_{\stm=\eps^*(\st)}\norm\st_\linf
 =\inf_{\stm=\eps^*(\st)}\braces{
    \ess\sup_{i,m,\pis\in\gO}\abs{\st_{im}(\pis)}
  }.
\]

\subsection{$\mu$-integrable distortions and equilibrated forces on bodies}
Following \cite{SCF} we use
$\lomu(\cl\gO,\rthree)$ to denote the space
of integrable vector fields on $\cl\gO$
whose restrictions to $\partial\gO$ are
integrable relative to the area measure on
$\partial\gO$.  On this space we use the
norm
\[
\norm\vf_{\lomu}=\ino\gO\abs{\vf}\,\dV
       +\ino{\partial\gO}\abs{\vf}\,\dA
        =\norm\vf_{\lone(\gO,\rthree)}+\norm\vf_{\lone(\partial\gO,\rthree)}.
\]
Alternatively, the $\lomu$-norm may be
regarded as the $\lone$-norm relative to the
Radon measure $\mu$, defined above
and hence the notation.

Forces, being elements of the dual space
$\lomu(\cloo,\rthree)^*$, may be identified
with elements of $\limu(\cloo,\rthree)$. A
force $\fc$ on a body, given in terms of a
body force $\bfc$ and a surface force
$\sfc$, may be identified with a continuous
linear functional relative to the
$\lomu$-norm if the body force components
$\bfc_i$ and surface force components
$\sfc_i$ (alternatively, $\abs{\bfc}$ and
$\abs{\sfc}$) are essentially bounded
relative to the volume and area measures,
respectively. In this case, the
representation is of the form
\[
\fc(\vf)=\ino\gO\bfc_i\vf_i\,\dV
    +\ino{\partial\gO}\sfc_i\vf_i\,\dA.
\]
Moreover, the dual norm of a force is the
$\limu$-norm, given as
\[
\norm\fc_\limu=\norm\fc^*_{\lomu}
  ={\esssup_{{{x\in\gO},\,{y\in\bndo}}}
  \braces{\abs{\bfc(x)},
                |\sfc(y)|}}
\]
as anticipated.

It is well known that if $\fc$ is
equilibrated, i.e., $\fc\in\proj^*_0(G)$,
for some
$G\in\paren{\lomu(\cl\gO,\rthree)\qrigs}^*$,
then,
\[
\ino\gO\bfc\,\dV
    +\ino{\partial\gO}\sfc\,\dA=0,
 \quad\text{and}\quad
 \ino\gO\pis\times\bfc\,\dV
    +\ino{\partial\gO}\pis\times\sfc\,\dA=0.
\]

\subsection{$\ld$-forces represented by body
forces and surface forces}
 \label{BodSurForces}
Using the trace operator $\trace$, for each
$\vf\in\ldo$ we may define
\[
\extnd(\vf)\from\cl\gO\to\rthree
\]
by
\( \extnd(\vf)(\pis)=\vf(\pis) \) for
$\pis\in\gO$ and
$\extnd(\vf)(y)=\trace(\vf)(y)$ for
$y\in\partial\gO$.  The trace theorem
\ref{Trace} and the original definition in
\ref{Definition} of the norm on $\ldo$ imply
that we defined a linear and continuous
mapping
\[
\extnd\from\ldo\tto\lomu(\cl\gO,\rthree).
\]
By the linearity of the extension mapping
and using $\extnd(\rig)=\rig$ for
$\rig\in\rigs$, we set
\[
\extnd\qrigs\from\ldo\qrigs
        \tto\lomu(\cl\gO,\rthree)\qrigs
\]
by $\extnd\qrigs([\vf])=[\extnd(\vf)]$. Thus
we have the following commutative diagram
\[
\begin{CD}
                 \lomu(\cl\gO,\rthree)
                            @<\extnd<< \ldo
                               \\
 {}         @V{\proj}VV       {}   @VV{\proj}V\\
                 \lomu(\cl\gO,\rthree)\qrigs
                   @<\extnd\qrigs<<
                               \ldo\qrigs.
\end{CD}
\]
The dual commutative diagram is
\[
\begin{CD}
                 \limu(\cl\gO,\rthree)
                            @>\extnd^*>>
                            \ldo^*
                               \\
 {}         @A{\proj^*}AA       {}   @AA{\proj^*}A\\
                 (\lomu(\cl\gO,\rthree)\qrigs)^*
                   @>(\extnd\qrigs)^*>>
                               (\ldo\qrigs)^*.
\end{CD}
\]
In particular, the image under $\extnd^*$ of
an equilibrated force
$\fc\in\limu(\cl\gO,\rthree)$ is an
equilibrated $\ld$-force.

As the norm of a mapping and its dual are
equal, we have%
\footnote{Note that we cannot use
    \[
    \norm{\extnd}
      =
         \norm{\extnd^*}=
              \sup_{\fc\in\limu(\cl\gO,\rthree)}
                \frac{\norm{\extnd^*(\fc)}}{\norm\fc^\limu}
      =
       \sup_\fc\frac{\inf_{\extnd^*(\fc)=\eps^*(\st)}
              \braces{\norm{\st}}}
            {\norm\fc^\limu}
    \]
    because $\eps^*$ is not surjective so there
    might be no $\st$ satisfying the condition
    $\extnd^*(\fc)=\eps^*(\st)$.}
\begin{align*}
\norm{\extnd\qrigs}
  &=
     \norm{(\extnd\qrigs)^*}=
          \sup_{G\in(\lomu(\cl\gO,\rthree)\qrigs)^*}
            \frac{\norm{(\extnd\qrigs)^*(G)}}{\norm{G}}
  \\&=
       \sup_{G\in(\lomu(\cl\gO,\rthree)\qrigs)^*}
            \frac{\inf_{(\extnd\qrigs)^*(G)=(\eps\qrigs)^*(\st)}\norm{\st}}{\norm{G}}
\end{align*}

Using the fact that the two mappings
$\proj^*$ are isometric injections onto the
respective subspaces of equilibrated forces,
we may replace $G$ above by an equilibrated
force $\fc\in\limu(\cl\gO,\rthree)$, and
$(\extnd\qrigs)^*(G)=(\eps\qrigs)^*(\st)$ is
replaced by $\delta^*(\fc)=\eps^*(\st)$.
Thus, we obtain
\[
\norm{\extnd\qrigs}=\sup_{\fc}
    \frac{\inf_{\delta^*(\fc)=\eps^*(\st)}
         \braces{\ess\sup_{i,m,\pis}\braces{\abs{\st_{im}(\pis)}}}}
           {\ess\sup_{i,x,y}\braces{\abs{\bfc_i(x)},\abs{\sfc_i(y)}}}\,,
\]
over all equilibrated forces in
$\limu(\cl\gO,\rthree)$.  Explicitly, the
condition $\delta^*(\fc)=\eps^*(\st)$ is
\[
\ino\gO\bfc\cdot\vf\,\dV+\ino\bndb\sfc\cdot\vf\,\dA
             =\ino\gO\st\cdot\eps(\vf)\,\dV
\]
as anticipated and we conclude that
\[
K=\norm{\delta\qrigs}.
\]
\begin{remk}
If we want to regard $\delta\qrigs$ as a
mapping between function spaces we should
use the decompositions of the respective
spaces into Whitney sums.  We already noted
that $\ldo\qrigs$ is isometrically
isomorphic to $\ldo_0$---the space of
$\ld$-vector fields having zero rigid
components.  Now $\lomu(\cl\gO,\rthree)_0$
is bijective to
$\lomu(\cl\gO,\rthree)\qrigs$ but as a
subspace of $\lomu(\cl\gO,\rthree)$ it has a
different norm (see Paragraph
\ref{Distortions}). Since we are interested
in the quotient norm in order to use the
essential supremum for the dual norm, we
will endow $\lomu(\cl\gO,\rthree)_0$ with
the quotient norm
$\norm{\vf_0}=\inf_{\rig\in\rigs}\norm{\vf_0-r}_\lomu$---
which brings us back to the problem of best
approximation by rigid velocity as described
in the end of paragraph \ref{Approx}.  Thus,
$\delta\qrigs$ becomes identical to the
restriction
\[
\delta_0=\delta\resto{\ldo_0}\from
    \ldo_0\tto\lomu(\cl\gO,\rthree)_0
\]
of $\delta$ to vector fields having zero
rigid components.  Its norm is given by
\[
\norm{\delta\qrigs}=\norm{\delta_0}
  =\sup_{\vf_0\in\ldo_0}\frac
  {\inf_{\rig\in\rigs}\braces{
     \displaystyle\int_\gO\sum_i\abs{\vf_{0i}-\rig_i}\,\dV
             +\displaystyle{\int}_{\!\!\!\partial\gO}\sum_i\abs{\vf_{0i}-\rig_i}\,\dA}}
     {\half\displaystyle\int_\gO\sum_{i,m}\abs{\vf_{0i,m}+\vf_{0m,i}}\,\dV}\,.
\]
Again, one may use smooth vector fields to
evaluate the supremum as these are dense in
$\ldo$.

\end{remk}

%
%
%
%
%
%
%
%
%
%
%

%

\appendix

\section{Elementary Properties of Quotient Spaces}
 \label{AppQuot}
We describe below some elementary properties
of quotient spaces of normed spaces (e.g.,
\cite[p.~227]{Taylor}).
\subsection{The quotient norm}
Let $\vs$ be a normed vector space with a
norm $\norm\cdot$ and $\rigs$ a closed
subspace of $\vs$ (e.g., a finite
dimensional subspace).  Then, the quotient
norm $\norm\cdot_0$ is defined on
$\vs\by\rigs$ by
\[
\norm{\vf_0}=\inf_{\vf\in\vf_0}\norm\vf.
\]
Denoting by $\proj\from\vs\tto\vs\by\rigs$
the natural linear projection
$\proj(\vf)=[\vf]$, we clearly have
$$
\norm{\proj(\vf)}_0=\norm{\proj(\vf+\rig)}_0
  =\inf_{\rig\in\rigs}\norm{\vf+\rig},
$$
 for any $\rig\in\rigs$.  The quotient norm
makes the projection mapping $\proj$
continuous and the topology it generates on
the quotient space is equivalent to quotient
topology.
\subsection{Dual spaces}
We note that as the projection $\proj$ is
surjective, its dual mapping
\[
\proj^*\from(\vs\by\rigs)^*\tto\vs^*
\]
is injective.  Clearly, it is linear and
continuous relative to the dual norms. If
$\phi\in\image\proj^*$ so
$\phi=\proj^*(\phi_0)$,
$\phi_0\in(\vs\by\rigs)^*$, then, for each
$\rig\in\rigs$,
\begin{align*}
\phi(\rig)&=\proj^*(\phi_0)(\rig)\\
           &=\phi_0(\proj(r))\\
           &=\phi_0(0)\\
           &=0.
\end{align*}

On the other hand, if for $\phi\in\vs^*$,
$\phi(\rig)=0$ for all $\rig\in\rigs$, then,
we may define $\phi_0\in(\vs\by\rigs)^*$ by
$\phi_0(\vf_0)=\phi(\vf)$, for some
$\vf\in\vs$ such that $\proj(\vf)=\vf_0$.
The choice of $\vf\in\vf_0$ is immaterial
because
$\phi(\vf+\rig)=\phi(\vf)+\phi(\rig)=\phi(\vf)$,
for any $\rig\in\rigs$.  We conclude that
$$
\image\proj^*=\rigs^\perp
 =\{\phi\in\vs^*\,|\,\phi(\rig)=0
   \text{ for all }\rig\in\rigs\}.
$$

Next we consider the dual norm of elements
of the dual to the quotient space.  For
$\phi_0\in(\vs\by\rigs)^*$, we have
\[
\norm{\phi_0}=\sup_{\vf_0\in\vs\by\rigs}
 \frac{\abs{\phi_0(\vf_0)}}{\norm{\vf_0}_0}.
\]

Thus,
\begin{align*}
\norm{\phi_0}&=\sup_{\vf_0\in\vs\by\rigs}
 \left\{\left.
 \frac{\abs{\phi_0(\proj(\vf))}}
  {\inf_{\rig\in\rigs}\norm{\vf+\rig}}\,\right|
  \text{ for some }\vf\in\vf_0
  \right\}\\
  &=
  \sup_{\vf_0\in\vs\by\rigs}
 \left\{\left.\sup_{\rig\in\rigs}
 \frac{\abs{\proj^*(\phi_0)(\vf)}}
  {\norm{\vf+\rig}}\,\right|
  \text{ for some }\vf\in\vf_0
  \right\}\\
  &=
  \sup_{\vf_0\in\vs\by\rigs}
 \left\{\left.\sup_{\rig\in\rigs}
 \frac{\abs{\proj^*(\phi_0)(\vf+\rig)}}
  {\norm{\vf+\rig}}\,\right|
  \text{ for some }\vf\in\vf_0
  \right\}\\
  &=
  \sup_{\vf_0\in\vs\by\rigs}
 \left\{\sup_{\vf'\in\vf_0}
 \frac{\abs{\proj^*(\phi_0)(\vf')}}
  {\norm{\vf'}}
  \right\}\\
  &=
  \sup_{\vf'\in\vs}
 \frac{\abs{\proj^*(\phi_0)(\vf')}}
  {\norm{\vf'}}
  \\
  &=\norm{\proj^*(\phi_0)}.
\end{align*}
We conclude that $\proj^*$ is norm
preserving.
%
%
%
%
%
%
%
%
%
%
\begin{ack}
  The research leading to this paper was
  partially supported by the Paul Ivanier
  Center for Robotics Research and Production
  Management at Ben-Gurion University.
\end{ack}



\end{document}